# *Ab initio* studies of electronic structure of defects in PbTe

Salameh Ahmad,[1] S. D. Mahanti,[1,*] Khang Hoang,[1] and M. G. Kanatzidis[2]

[1]Department of Physics and Astronomy, Michigan State University, East Lansing, MI 48824, USA

[2]Department of Chemistry, Michigan State University, East Lansing, MI 48824, USA

[*]Corresponding author. Electronic mail: mahanti@pa.msu.edu

**Abstract**

Understanding the detailed electronic structure of deep defect states in narrow band-gap semiconductors has been a challenging problem. Recently, self-consistent *ab initio* calculations within density functional theory (DFT) using supercell models have been successful in tackling this problem. In this paper, we carry out such calculations in PbTe, a well-known narrow band-gap semiconductor, for a large class of defects: cationic and anionic substitutional impurities of different valence, and cationic and anionic vacancies. For the cationic defects, we study a series of compounds $R\text{Pb}_{2n-1}\text{Te}_{2n}$, where $R$ is vacancy or monovalent, divalent, or trivalent atom; for the anionic defects, we study compounds $M\text{Pb}_{2n}\text{Te}_{2n-1}$, where $M$ is vacancy, S, Se or I. We find that the density of states (DOS) near the top of the valence band and the bottom of the conduction band get significantly modified for most of these defects. This suggests that the transport properties of PbTe in the presence of impurities can not be interpreted by simple carrier doping concepts, confirming such ideas developed from qualitative and semi-quantitative arguments.

## I. INTRODUCTION

Narrow band-gap IV-VI binary semiconductors and their alloys have been of great interest because of their fundamental electronic properties and for their practical applications. Lead chalcogenide salts PbTe and PbSe are two IV-VI narrow gap semiconductors whose studies over the last several decades have been motivated by their importance in infrared detectors, light-emitting devices, infrared lasers, thermophotovoltaics and thermoelectrics.[1-3] In fact, PbTe was one of the first materials studied by Ioffe and his colleagues in the middle of the last century when there was a revival of interest in thermoelectricity.[4] This compound, its alloys with SnTe and PbSe, and related compounds called TAGS (alloys of $AgSbTe_2$ and GeTe) were for many years the best thermoelectric materials at temperatures ~700 K.[5] In recent years, quantum wells of $PbTe/Pb_{1-x}Eu_xTe$, $PbSe_{0.98}Te_{0.02}/PbTe$ superlattices[6] and novel quaternary compounds $AgSbPb_{2n-2}Te_{2n}$ (n=9,10)[7] have attracted considerable attention because of their large thermoelectric figure of merit (FOM). Similar quaternary systems with Ag replaced by Na are also promising high temperature thermoelectrics.[8] Most of the above systems have stoichiometry closer to the parent compound PbTe.

It is well-known that transport and optical properties of semiconductors are dominated by the states in the neighborhood of the band gap. For example, in thermoelectrics, the FOM (denoted as $ZT$ where $T$ is the operating temperature) depends on the thermopower $S$ and electric conductivity $\sigma$ through the relation $ZT = \sigma S^2 T/\kappa$, where $\kappa$ is the sum of electronic and lattice thermal conductivities of the material.[9] Clearly, large values of $ZT$ require large values of $S$ and $\sigma$ both of which depend sensitively on the nature of the electronic states near the band gap.[10] Thus it is important to have a fundamental understanding of the changes in the electronic states near the band-gap region in PbTe when it is mixed with other binary and ternary compounds. Before being able to understand the nature of the electronic states near the band gap and/or the Fermi energy in the complex systems mentioned above, one must understand how the electronic states of PbTe change when Pb or Te is replaced by impurity atom or by vacancy.

A detailed microscopic understanding of the nature of the defect states in PbTe is an old but a challenging problem. Naively, one expects to see bound shallow defect states in the band gap (pinned to the valence band maximum or conduction band minimum) when a Pb or a Te atom is replaced by either a donor- or acceptor-like impurity or by a vacancy. However, it is known that, in narrow band-gap semiconductors like PbTe, one does not see these shallow defect states but deep defect states which can be outside the band gap.[11-15] Theoretically, a fundamental understanding of the shallow defect states in semiconductors with large band gap traces back to the classic work of Kohn and Luttinger.[16] In this case, bound defect states are produced by the long-range Coulomb potential of charged defects screened by the dielectric constant of the host semiconductor. These states appear below the CBM (for donor impurities) or above the VBM (for acceptor impurities). Their binding energies are well-described by a hydrogenic model after taking proper account of the band effective mass, dielectric screening, and a small central-cell correction.[12] However, even in large band-gap semiconductors, one can have deep defect states (strongly bound) when the central-cell effects are equally important as the one caused by the Coulomb potential.[12] Unlike the shallow impurity levels, the deep levels (in narrow band-gap semiconductors) are dominated by the short-range atomic-like defect potential and changes in local bonding. Consequently, the detailed physics underlying the formation of deep defects states in these systems depends sensitively on the specific features of the defect.

One of the earlier attempts to understand the origin of deep defect states in PbTe was made by Lent *et al.*[13] For substitutional defects in PbTe, they presented a simple chemical theory for a large class of substitutional defects and vacancies in terms of the atomic energy levels and tight binding concepts. Unfortunately, these calculations did not take into account the potentials generated by the impurities in a self-consistent manner and, therefore, were only qualitative in their prediction. In fact, an even earlier attempt to understand the origin of vacancy-induced deep defect states in PbTe was made by Parada and Pratt.[14] They used a Slater-Koster model and Wannier function basis constructed out of a finite number of PbTe bands. They predicted defect states outside the band gap and were able to explain the low temperature *n*- and *p*-type transport properties in nominally pure PbTe. The defect potential used in the Slater-Koster model was assumed to be a constant and was not calculated self-consistently. Therefore the precise positions of the defect states predicted by this calculation are not expected to be quantitatively reliable. The other major efforts in understanding the physics of deep defect states are for In-doped PbTe. Inhomogeneous mixed-valence models for In and other group III impurities have been proposed to explain several experimental data.[17] In these models, In impurities exist in two valence states, one trivalent and the other monovalent, stabilized by strong electronic and ionic relaxations. Our recent calculations have cast some doubts on the validity of this model.[18]

In this paper, we discuss the results of our attempts to understand the physics behind the deep defect states in a large class of defects in PbTe using self-consistent *ab initio* electronic structure calculations within density functional theory (DFT). Only a few *ab initio* studies have been made recently. One by Bilc *et al.*[19] who have recently reported their results on Ag and Sb substitutional defects in PbTe using all-electron full-potential linearized augmented planewave (FP-LAPW) method.[20] Another by Ahmad *et al.*[18] who looked at the In impurity problem. Similar calculations have been carried out by Edwards *et al.*[21] who looked at the Ge and Te vacancy and anti-site defects in GeTe, and by Lany *et al.*[22] on the DX centers in CdTe. The purpose of our present work is to extend the previous studies of the defects in PbTe to many other different defects: Pb and Te vacancies, Pb substituted by monovalent atoms (alkalis, Cu), divalent atoms (*s*-type Zn, Cd, Hg and *p*-type Ge, Sn), or trivalent atoms (group III Ga, In, Tl, and group V As, Sb, Bi), and Te substituted by S, Se, or I. Our main focus is to develop a systematic understanding of these defect states as one goes across or down the periodic table to select different impurity atoms. Preliminary results of this study have been recently reported.[23,24] The arrangement of this paper is as follows. In Sec. II, we describe the structural model and discuss briefly the method used to calculate the electronic structure. In Sec. III, we present our calculation results focusing on the electronic density of states (DOS) with and without the defects. We give the change in the total and partial DOS caused by different types of defects and discuss the underlying physics. In Sec. IV, we discuss our theoretical results in the light of recent measurements of thermopower in several PbTe based compounds. Finally, we give a brief summary in Sec. V.

## II. METHOD OF CALCULATION

Before we discuss the structural model used to study the impurity states, we would like to briefly describe the quaternary systems one constructs in developing novel thermoelectrics starting from the binary compound PbTe and replacing Pb atoms by other atoms. The new compounds are described by the formula $RR'Pb_{2n-2}Te_{2n}$, where $R$ and $R'$ are chosen to maintain charge neutrality. Here two divalent Pb atoms are replaced by one monovalent ($R$) and one trivalent ($R'$) atom. If $R$ is divalent then $R'$ is also divalent (for example $R=R'=$Cd). The material is therefore charge compensated and most likely will maintain its semiconducting character. One then adjusts the concentration of $R$ and $R'$ to control the sign of the charge carriers and optimize their concentration. In a recent work, Bilc *et al.*[19] have discussed the electronic structures of a series of charge compensated $AgSbPb_{2n-2}Te_{2n}$ compounds (n=6, 9, 16) and also charge uncompensated compounds $AgPb_{2n-1}Te_{2n}$, $SbPb_{2n-1}Te_{2n}$ (n=16) where only one of the elements $R$ or $R'$ is present. They found resonant states overlapping conduction and valence bands caused by Sb and Ag impurities, respectively. In this paper, we focus on defects of one type (either R or R') and investigate the electronic structure of $R$Pb$_{2n-1}$Te$_{2n}$. We also carry out similar calculations in compounds of the type $M$Pb$_{2n}$Te$_{2n-1}$, where $M$ is a defect at the Te site. Our aim is to explore systematically the impurity-induced changes in the electronic DOS as one goes down or across the periodic table.

We use a supercell model where the defects are periodically arranged in a PbTe lattice. This corresponds to the formula $R$Pb$_{2n-1}$Te$_{2n}$ or $M$Pb$_{2n}$Te$_{2n-1}$ as mentioned above. To increase the distance between the defects as much as possible (in order to reduce the interaction between them) without increasing the size of the supercell enormously we have chosen n=16. This corresponds to a unit cell containing 64 atoms with the minimum distance between two defects being ~13Å (see Fig. 1). Due to the large dielectric constant of the host PbTe, we expect long-range Coulomb effects to be strongly screened, thereby reducing the impurity-impurity interaction. However, short-range interactions are still present. We will address this issue when we discuss the partial DOS associated with different atoms.

Electronic structure calculations have been performed within DFT using all-electron FP-LAPW plus local orbital method[20] incorporated in WIEN2k.[25] We have used the generalized-gradient approximation (GGA)[26] for the exchange and correlation potential. Scalar relativistic corrections were included and spin-orbit interaction (SOI) was implemented using a second variational procedure.[27] Convergence of the self-consistent iterations was performed using 20 $k$-points inside the reduced Brillouin zone to within 0.0001 Ry (1.36 meV) with a cutoff energy of -6.0 Ry separating the valence and the core states. For this initial survey, calculations have been performed using the experimental lattice constant of PbTe, which is 6.454Å,[28] and no relaxation (volume or internal) studies have been carried out. For a few systems, we have also used projector augmented-wave (PAW)[29] method incorporated in *Vienna Ab-initio Simulation Package* (VASP)[30] to compare with the FP-LAPW results, increase the impurities separation by going to larger supercells. In our future work, we propose to carry out complete relaxation studies in all the defect systems.

### III. RESULTS AND DISCUSSION

Before presenting the results of the *ab initio* electronic structure calculations, we discuss some basic physics of PbTe system. In this compound Pb and Te 6$s$ and 5$s$ states are quite deep and the valence and conduction bands are formed primarily out of 6$p$ states of Pb and 5$p$ states of Te. The Pb-Te bonds have both covalent and ionic character. The removal of one Pb atom or replacing it by a monovalent cation leads to a local charge disturbance and alters the covalent bonding with the neighboring Te (denoted as Te2) $p$ states. As a result, the valence band states get strongly perturbed. One expects acceptor-like states to appear near the top of the valence band. However, their precise positions can only be obtained after a detailed self-consistent calculation. Also, because of the local nature of the perturbation we expect the entire valence band region to be affected. In case of the monovalent Ag impurity, one has, in addition, the filled $d$ states which interact with the Te $p$ states and change the valence band. The Cd impurity, although divalent like Pb, has 5$s$ state which may give rise to new states near the bottom of the conduction band. Indium (In) and other trivalent $sp$ atoms may also give rise to a bound $s$ state below the valence band and therefore act more like a monovalent impurity than a trivalent one. It is important to emphasize that simple shallow impurity-state models[16] or Slater-Koster type models without calculating the impurity potential self-consistently[13-15,17] are not capable of describing the short-range physics of these impurity states quantitatively. Only a careful self-consistent *ab initio* calculation will be able to do so.

### A. Pb-site defects

The removal of one Pb atom (i.e., creation of one vacancy at the Pb site) or replacing it by another atom with different valence introduces a local charge disturbance and alters the bonding with the neighboring Te $p$ states. Due to the long-range nature of the former and the short-range of the latter, we expect the entire valence and conduction band states to get perturbed in the presence of the impurity.

#### *1. Pb vacancy in PbTe*

The total DOS of PbTe with and without Pb vacancy and the difference between the two are shown in Fig. 2(a)-(c). For the sake of comparison, we shifted the DOS of PbTe by a small amount (~0.2 eV) so that the lower edge of the valence-band DOS of the systems with and without Pb vacancy matched. We found that the first peak below the valence band maximum (VBM) also matched after this shift. This small shift appears to be reasonable since with this shift the Te core 5$s$ bands also matched perfectly. As we can see from Fig. 2(a)-(c), the entire valence-band and conduction-band DOS get modified by the presence of the Pb vacancy. It is important to note that the states near the bottom of both the valence and conduction band do not change appreciably in the presence of the Pb vacancy. Since transport properties are controlled by the states near the VBM and conduction band minimum (CBM), let us focus on the band gap region within an energy range of ± 0.5 eV about the center of the band gap. We see that there is very little change near the bottom of the conduction-band DOS whereas there is an increase in the DOS near the top of the valence band. We analyze this increase in more detail below. From Fig. 2(a), we can clearly see that the Pb vacancies give rise to holes as expected.

To see which atoms contribute to the change in the DOS near the VBM, we show in Fig. 2(b) the difference in the partial DOS between PbTe with and without Pb vacancy. There are six nearest-neighbor Te atoms of the Pb vacancy (denoted as Te2) and three other inequivalent Te atoms (denoted as Te1, Te3, and Te4). Among the Pb neighbors of the Pb vacancy, Pb2 is the Pb atom that bridges two Te2 atoms each being the nearest-neighbor of one of the defect sites. In one supercell there are three Pb2 atoms and twenty-eight other Pb atoms. Of the latter, there are four inequivalent (Pb3, Pb4, Pb5, Pb6) lead atoms. Clearly the major contribution to the DOS increase near the top of the valence band comes from the six nearest-neighbor Te2 atoms and we see that Te2 states in the lower half of the valence band get pushed towards the upper half. This transfer of the Te2 DOS is partially compensated by the downward shift of the DOS associated with the other atoms. Also other Te atoms do contribute a small amount to the increased DOS near the VBM.

It should be noted that, in FP-LAPW calculations, there is an appreciable DOS associated with charges outside the atomic spheres, i.e., in the interstitial space. In Fig. 2(c), we show the difference in the total DOS ($\Delta\rho_{tot}$), in the DOS associated with the charges inside the atomic spheres ($\Delta\rho_{at}$), and in the DOS associated with the interstitial space ($\Delta\rho_i=\Delta\rho_{tot}-\Delta\rho_{at}$). $\Delta\rho_i$ and $\Delta\rho_{at}$ have almost the same energy dependence and are comparable in magnitude. We can interpret the peak in $\Delta\rho_{tot}$ near the top of the valence band [Fig.2(c)] as a vacancy-induced resonant state. This should be significant in the hole-doped samples where these states will be involved in charge and energy transport. The fact that Pb2, Te1, Te3, and Te4 contribute very little to $\Delta\rho_{tot}$ suggests that the vacancy-induced resonant states are weakly interacting.

It is interesting to compare our results with those of Parada and Pratt.[14] They found that the perturbation due to a Pb vacancy was not strong enough to drive any levels out of the Te $p$ bands. The Pb $s$ state associated with the vacancy site was driven about 3 eV above the bottom of the conduction band, hence of no significance for transport. A doubly degenerate state was driven out of the Te $s$ band but by about 1 eV, hence it was occupied by two Te $s$ electrons. As a result, a Pb vacancy creates two holes in the Te $p$ valence band and leads to $p$-type conduction. These findings are supported by our *ab initio* calculations. However, the subtle effects such as the DOS enhancement near the top of the Te *p* valence band which can affect the thermopower of holes was not investigated by them.

### 2. Monovalent impurities (Na, K, Rb, Cs, Cu, Ag) in PbTe

Now let us discuss the monovalent impurities. They should provide a weaker perturbation compared to the vacancy. In Fig. 3, instead of giving the total DOS, we only show the difference ($\Delta\rho_{tot}$). We find that K, Rb, Cs, Cu and Ag all show a peak near the top of the valence band, just like a vacancy. Na is anomalous, it does not seem to change the DOS near the valence band top, but shows an increase (as compared to other three alkalis) near -3.2 eV. The major difference between Cu, Ag and the alkalis is seen near the bottom of the valence band in the energy range (-2, -1) eV for Cu and (-4, -3) eV for Ag. This is due to the Cu and Ag *d* states hybridizing with the Te *p* bands. We will further discuss these results later in Sec. IV when we compare our theoretical results with experiment. We find that the valence *s* state of the alkalis are all above the valence band, hence all of them donate one electron to the system. In contrast to the suggestions made in Ref.13, we find that Cs behaves just like Na in this respect. We should note that even if the alkali atoms donate one electron when they replace Pb (which is divalent), they give rise to one hole per impurity and hence act as acceptors.

### 3. Divalent impurity (Zn, Cd, Hg, Sn, Ge) in PbTe

There are two types of divalent impurities: *s*-type (Zn, Cd, and Hg) and *p*-type (Sn and Ge). The valence configurations of the *s*-type impurities are $4s^2$ (Zn), $5s^2$ (Cd) and $6s^2$ (Hg); whereas those for the *p*-types are $4p^2$ (Sn) and $5p^2$ (Ge). The valence state of divalent Pb is of course $6p^2$. We find that the *s*-type impurities introduce resonant states near the bottom of the PbTe conduction band and there are three major changes in the DOS: (i) an enhancement of the DOS near the bottom of the conduction band caused by the resonant state, (ii) a reduction in the DOS near the top of the valence band within an energy range ~0.25 eV, and (iii) an enhancement of the DOS near the bottom of the valence band [see Fig. 4 and Fig. 5(a)-(d)]. These three changes can be seen clearly if, for example, one looks at the Cd *s* character in Fig. 5(d). The Cd atom introduces an *s*-like state near the bottom of the conduction band. This state strongly hybridizes with the *p* orbitals of the six neighboring Te2 atoms and the bonding state appears at about -4 eV [see Fig. 5(c) and 5(d)]. The corresponding anti-bonding state appears as a resonant state near the conduction band bottom, reflecting the deep defect nature of these impurity states. The states near the top of the valence band (anti-bonding between Pb and Te2 *p*'s) are depressed when Pb is replaced by Cd. Because of the resonant state near the bottom of the conduction band, we expect electron-doped Cd systems to exhibit large thermopower. Its experimental implication will be discussed later. In contrast to the *s*-type divalent impurities, the *p*-type impurities, Sn and Ge, do not alter the DOS near the band gap (see Fig. 6). They behave very much like Pb near the band gap region, which is understandable since Pb is also a *p*-type divalent atom.

### 4. Trivalent impurity (Ga, In, Tl, As, Sb, Bi) in PbTe

Like the divalent case, we also have two types of trivalent impurities: group III (Ga, In, Tl) and group V (As, Sb, Bi). The valence configurations of the group III trivalent atoms are $4s^2 4p^1$, $5s^2 5p^1$ and $6s^2 6p^1$ for Ga, In and Tl, respectively; whereas the valence configurations of the group V trivalent atoms are $4p^3$, $5p^3$ and $6p^3$ for As, Sb and Bi, respectively. All three group III impurities show a strongly bound state below the bottom of the valence band and a state near the band-gap region. These strongly bound states can be identified with the electrically inactive "hyperdeep" levels proposed by Hjalmarson *et al.*[31] which always appear along with the electronically active deep defect states in the band-gap region. These deep defect states are seen in our calculation as peaks in $\Delta\rho_{tot}$ as shown in Fig. 7 near the Fermi energy (at 0 eV).

Results for Ga and Tl atoms show that their deep defect states are resonant states near the top of the PbTe valence band (Fig. 7), whereas for In this state lies in the band-gap region.[18] This difference between In and the other two may explain the experimental anomalies seen in the case of In impurities in PbTe.[17] The resonant states at the top of the valence band for the Ga and Tl is caused by the hybridization of the *s* states of these atoms with the *p* state of Te2. We show, in Fig. 8(a)-(d), the total and partial DOS of PbTe with and without Ga impurity, the difference $\Delta\rho_{tot}$, and *s* character of the partial DOS of Ga impurity. Results for Tl impurity are similar excepting that the "hyperdeep" level is further below the valence band minimum. The "hyperdeep" level at about -5.0 eV is predominantly impurity(Ga)-like whereas the resonant state centered around 0 eV is predominantly host(Te2)-like, which is consistent with the picture proposed by Hjalmarson *et al.*[31] The peak of the resonant state is about 0.25 eV below the VBM. This is consistent with the current understanding of the Tl-doped PbTe systems (see Ref. 17 and references therein). According to our calculations, Ga and Tl impurities should show rather similar behavior.

Indium (In) is different from Ga and Tl. The deep defect state lies in the band gap and hence localized. This localized state is occupied by one electron. According to the mixed-valence model of In impurities in PbTe,[17] two In impurities substituting for two Pb atoms give 2 electrons each to the

Te $p$ band leaving them as two In$^{2+}$ ions. However, a pair of In$^{2+}$ ions undergo charge disproportionation and become In$^{3+}$ and In$^{1+}$ resulting in an inhomogeneous mixed-valence state for the pair. In our calculations, we see that to create an In$^{3+}$ ion will require taking two electrons from the "hyperdeep" levels to the top of the valence band and will cost too much energy. We have checked this by taking two In impurities in our 64 atom unit cell and allowing for lattice relaxation. We do not find any evidence for the presence of an In$^{3+}$ state. [18]

Results for group V impurities (As, Sb and Bi) show that they introduce resonant states near the bottom of the conduction band (see Fig. 9). In contrast to the group III impurities which have $s$ and $p$ valence electrons, the resonant states for these $p$ valence electron impurities are rather broad, the sharpest structure being for As. In Fig. 10(a)-(d), we show the total and partial DOS of PbTe with the As impurity. The resonant state is a mixture of the $p$ states of As and the $p$ states of the neighboring Te2 atoms in almost equal proportions.

### B. Te-site defects

As in the case of Pb-site defects, the removal of one Te atom (i.e., creation of one vacancy) or replacing it by other atom introduces a local charge disturbance and alters the covalent bonding with the neighboring Pb (denoted as Pb2) $p$ states. As a result, we expect both the valence and conduction bands states to get strongly perturbed.

#### *1. Te vacancy in PbTe*

The total and partial DOS of PbTe with and without vacancy and the difference $\Delta\rho_{tot}$ are shown in Fig.11(a)-(c). For the sake of comparison, we shifted the DOS of PbTe with defect by ~0.455 eV such that the core bands match perfectly. As we can see, the entire valence and conduction band DOS get perturbed by the vacancy. Since transport properties are controlled by the states near the Fermi energy (at 0 eV), let us focus on this region. As seen in Fig. 11(a)-(c), new states appear in the band gap region and near the bottom of the conduction band in the energy range -0.5 to 0 eV. Partial DOS analysis indicates that these states are predominantly of Pb $p$ character [see Fig. 11(c)]. The valence band looses 6 states (including spin) per Te vacancy. These 6 states are occupied by 4 Te electrons and 2 Pb electrons. Removal of Te atom takes 4 valence electrons away leaving two Pb valence electrons. As a result, two Pb electrons occupy the new states mentioned above. The system then behaves like an $n$-doped system. Our picture is qualitatively consistent with that of Parada[15] who argued that each Te vacancy causes 4 states (8 states including spin) to be removed from the valence band and to move near the conduction band bottom. This picture includes the Te 5s core state. Since Te atom has 6 electrons, the two electrons (coming from Pb) occupying the 8 states in valence band (in PbTe) have to move into the conduction band resulting in an *n*-type conduction. However, the detailed nature of the DOS near the bottom of the conduction band could not be obtained in earlier non-selfconsistent calculations.[15]

### 2. S and Se impurities in PbTe

S and Se, like Te, need two electrons to fill their valence *p* shell. Thus one expects mainly the valence band to be perturbed when one replaces Te by S or Se and there should be a small perturbation of the conduction band (indirectly through changes in the hybridization effects). This is clearly seen in Fig. 12(a)-(c) and Fig. 13(a)-(c). The change in the valence-band DOS takes place over the entire valence-band energy range with a reduction in the DOS in the energy range of 0.5 eV from the top of the valence band. S and Se *p* states appear in the entire valence-band region peaking around 1.2-1.5 eV below the VBM [Fig. 12(d) and Fig. 13(d)]. The conduction band is hardly affected within about 1.0 eV from its minimum.

### 3. I impurity in PbTe

Iodine (I) appears to act as an ideal donor in PbTe, in agreement with what is known about the halogen impurities in PbTe.[17,32] It does not perturb the states both near the top of the valence band and near the bottom of the conduction band as can be seen in Fig. 14(a)-(d). The major change in the DOS occurs near the bottom of the valence band around -5.0 eV (predominantly I *p* character), near -1.1 eV and +1.0 eV due to changes in the Pb *p* hybridization effect. The Fermi energy is shifted to the conduction band making it an *n*-type impurity (donor).

### IV. COMPARISON WITH EXPERIMENT

One of the remarkable properties of lead chalcogenide group of narrow band-gap semiconductors is that they have a range of non-stoichiometry accompanied by either Pb or Te vacancies. Our theoretical results predict that Pb vacancies produce *p*-type PbTe whereas Te vacancies produce *n*-type PbTe, which are consistent with the earlier model calculations.[14,15] However, in contrast to these earlier semi-empirical or model studies, *ab initio* calculations predict drastic changes in the DOS near the band gap region, which should affect the transport properties quantitatively.

As regards monovalent impurities, Poudeu *et al.*[8] found that $Na_{0.95}Pb_mSbTe_{m+2}$ (m=20) was a *p*-type semiconductor with high thermoelectric FOM, $ZT_{max}$ of ~ 1.7 at 650 K. The observed *p*-type behavior is consistent with our *ab initio* calculations and what is known about Na and Li impurities in PbTe. To approach a Na concentration closer to that for m=20 compound, we add one more Na atom at the middle of the 64 atom super cell. The calculation shows that the DOS in the region of the band gap does not change by this addition. Thus one can treat the $Na_{0.95}Pb_{20}SbTe_{22}$ as basically PbTe doped with holes. Its thermopower should be comparable to that of *p*-doped PbTe.

The behavior of trivalent impurities such as In, Ga, Tl has been both interesting and puzzling. Gelbstein *et al.*[33] have recently argued that the thermoelectric property of In-doped PbTe can be understood in terms of the existence of deep-lying states generated by In-doping. We find that In not only introduces deep defect states in the band gap but also strongly reduces the DOS near the top of the valence band. This results in the reduction in the minority (hole) contribution to thermopower both due to DOS reduction and annihilation of the holes by electrons occupying the localized state in the gap. The net effect can be an increased *n*-type thermopower.

As, Sb, and Bi give rise to resonant states near the bottom of the conduction band. Although this resonant state may not take part directly in transport properties, it indirectly affects the transport properties by increasing the DOS near the bottom of the conduction band. This was pointed out in Fig. 9 and Fig.10. The increased DOS can in principle give rise to an increased thermopower in *n*-type systems. A simple calculation[34] of thermopower by assuming energy-independent relaxation time showed that room temperature thermopower of $AgSbPb_{2n-2}Te_{2n}$ can be large, as seen experimetally.[7] However, detailed calculations of transport coefficients taking into account the changes in band structure (caused by the Sb induced resonant states) and energy dependent scattering indicates that this enhancement of thermopower is not true in general but depends on the carrier concentration and temperature range studied.[35]

### V. SUMMARY

In summary, our *ab initio* electronic structure calculations in $RPb_{2n-1}Te_{2n}$ (n=16), where *R* is a vacancy, or monovalent, divalent, or trivalent atom, show that when a Pb atom is substituted by *R*, the DOS gets perturbed over the entire valence and conduction bands. There are major changes in the DOS near the band gap region for most *R*. This should have significant impact on the transport properties of these new compounds. We find that Na does not change the DOS within 0.5 eV of the band maxima; thus it is an ideal acceptor. In contrast, other alkali atoms and Ag and Cu give rise to an increase in the DOS near the top of the valence band and negligible change in the DOS near the bottom of the conduction band. Hg, Cd, Zn give rise to strong resonant state near the bottom of the conduction band and suppress the DOS near the top of the valence band (which is good for *n*-type thermoelectrics). Group V (As, Sb and Bi) impurities introduce resonant states near the bottom of the conduction band and should be good for *n*-type thermoelectrics. In contrast, the trivalent impurities (Ga, In, and Tl) either introduce bound states in the gap or resonant states near the top of the valence band. Te vacancy also has a strong effect on the DOS of PbTe near the band gap region. New states appear in the band gap below the CBM, these states comprise primarily of the *p* states of Pb neighboring the vacancy. In contrast, the Pb vacancy increases the DOS near the top of the valence band but by a small amount. Te vacancy should be *n*-type whereas Pb vacancy should be *p*-type. Iodine appears to be an ideal donor, it does not change the DOS of PbTe near the band gap, just shifts the Fermi energy to the conduction band. The divalent atoms S and Se also do not change the DOS near the CBM. There is however some depletion of the DOS near the VBM. These results should have important implications in the thermoelectric properties of the *n*-type ternary compound PbTe$_{1-x}$S$_x$.[36] One can synthesize mixed Te/S systems without appreciably affecting the conduction-band states and hence the *n*-type charge and energy transport at the same time reduce the thermal conductivity through microstructure formation.

## ACKNOWLEDGMENT

This work was partly supported by the MURI grant (No. N00014-03-10789) from the Office of Naval Research. We acknowledge helpful discussions with Dr. P. Poudeu and Dr. J. Androulakis.

**Figure Captions**

Figure 1 (color online): Supercell model of $R$Pb$_{31}$Te$_{32}$ where $R$ is either a vacancy or an impurity atom. Small balls are for Pb and Te, large balls are for $R$.

Figure 2: (a) The total DOS of PbTe with and without Pb vacancy, (b) the difference in the partial DOS between PbTe with and without Pb vacancy where the six nearest neighbor Te atoms of the Pb vacancy are denoted as Te2 and there are three other inequivalent Te atoms (denoted as Te1, Te3, and Te4), and (c) the difference in the total DOS ($\Delta\rho_{tot}$), in the DOS associated with the charges inside the atomic spheres ($\Delta\rho_{at}$), and in the DOS associated with interstitial regions ($\Delta\rho_i = \Delta\rho_{tot} - \Delta\rho_{at}$). The energy origin in (b) and (c) is chosen to be the top of the valence band of PbTe; whereas the energy origin in (a) is the highest occupied Kohn-Sham single-particle state (which will be denoted as the Fermi energy) of the system with defect.

Figure 3: The difference in the total DOS ($\Delta\rho_{tot}$) between PbTe with and without monovalent impurities: Na, K, Rb, Cs, Cu, and Ag. The energy origin (0 eV) is chosen to be at the top of the valence band of PbTe.

Figure 4: The difference in the total DOS ($\Delta\rho_{tot}$) between PbTe with and without $s$-type divalent atom: Zn, Cd and Hg. The energy origin (0 eV) is chosen to be at the top of the valence band of PbTe.

Figure 5: (a) The total DOS of PbTe with and without Cd impurity, (b) their difference $\Delta\rho_{tot}$, (c) partial DOS per Te associated with nearest neighbor Te atoms, and (d) Cd partial DOS of s-character. The Fermi energy of PbTe with Cd impurity is at the energy origin (0 eV).

Figure 6: The difference in the total DOS ($\Delta\rho_{tot}$) between PbTe with and without $p$-type divalent atom: Sn and Ge. The energy origin (0 eV) is chosen to be at the top of the valence band of PbTe.

Figure 7: The difference in the total DOS ($\Delta\rho_{tot}$) between PbTe with and without trivalent group III impurities: Ga, In, and Tl. The energy origin (0 eV) is chosen to be at the top of the valence band of PbTe.

Figure 8: (a) The total DOS of PbTe with and without Ga impurity, (b) their difference $\Delta\rho_{tot}$, (c) partial DOS per Te associated with nearest neighbor Te atoms, and (d) Ga partial DOS of $s$ character. The Fermi energy of PbTe with Ga impurity is at the energy origin (0 eV).

Figure 9: The difference in the total DOS ($\Delta\rho_{tot}$) between PbTe with and without trivalent group V impurities: As, Sb, and Bi. The energy origin (0 eV) is chosen to be at the top of the valence band of PbTe.

Figure 10: (a) The total DOS of PbTe with and without As impurity, (b) their difference $\Delta\rho_{tot}$, (c) partial DOS per Te associated with nearest neighbor Te atoms, and (d) As partial DOS of $p$ character. The Fermi energy of PbTe with As impurity is at the energy origin (0 eV).

Figure 11: (a) The total DOS of PbTe with and without Te vacancy, (b) their difference $\Delta\rho_{tot}$, and (c) partial DOS per Pb associated with nearest neighbor Pb atoms. The Fermi energy of PbTe with Te vacancy is at the energy origin (0 eV).

Figure 12: (a) The total DOS of PbTe with and without S impurity, (b) their difference $\Delta\rho_{tot}$, (c) partial DOS per Pb associated with nearest neighbor Pb atoms, and (d) S partial DOS of $p$ character. The Fermi energy of PbTe with S impurity is at the energy origin (0 eV).

Figure 13: (a) The total DOS of PbTe with and without Se impurity, (b) their difference $\Delta\rho_{tot}$, (c) partial DOS per Pb associated with nearest neighbor Pb atoms , and (d) Se partial DOS of $p$ character. The Fermi energy of PbTe with Se impurity is at the energy origin (0 eV).

Figure 14: (a) The total DOS of PbTe with and without I impurity, (b) their difference $\Delta\rho_{tot}$, (c) partial DOS per Pb associated with nearest neighbor Pb atoms, (d) I partial DOS of $p$ character. The Fermi energy of PbTe with I impurity is at the energy origin (0 eV).

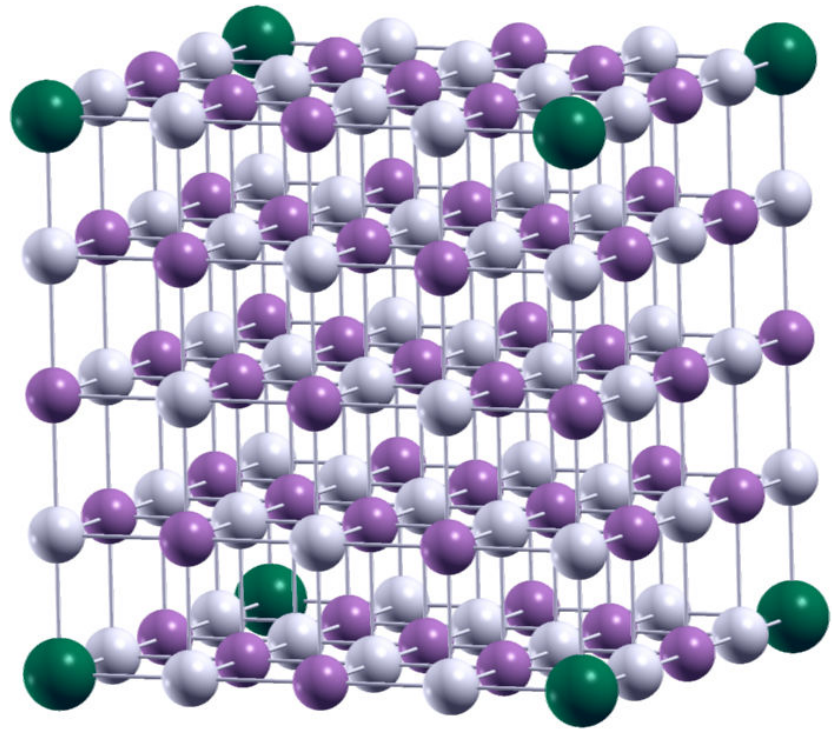

Figure 1

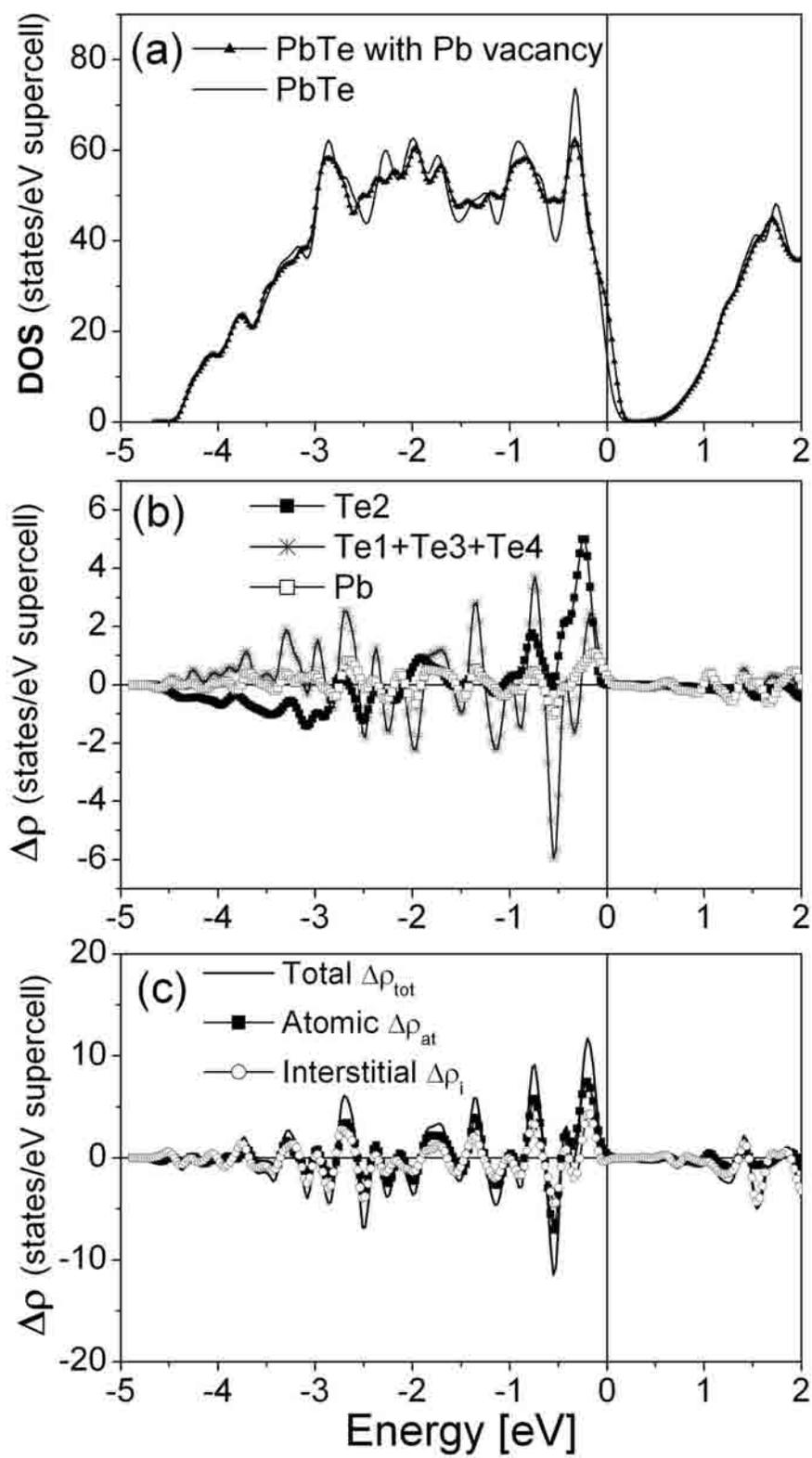

Figure 2

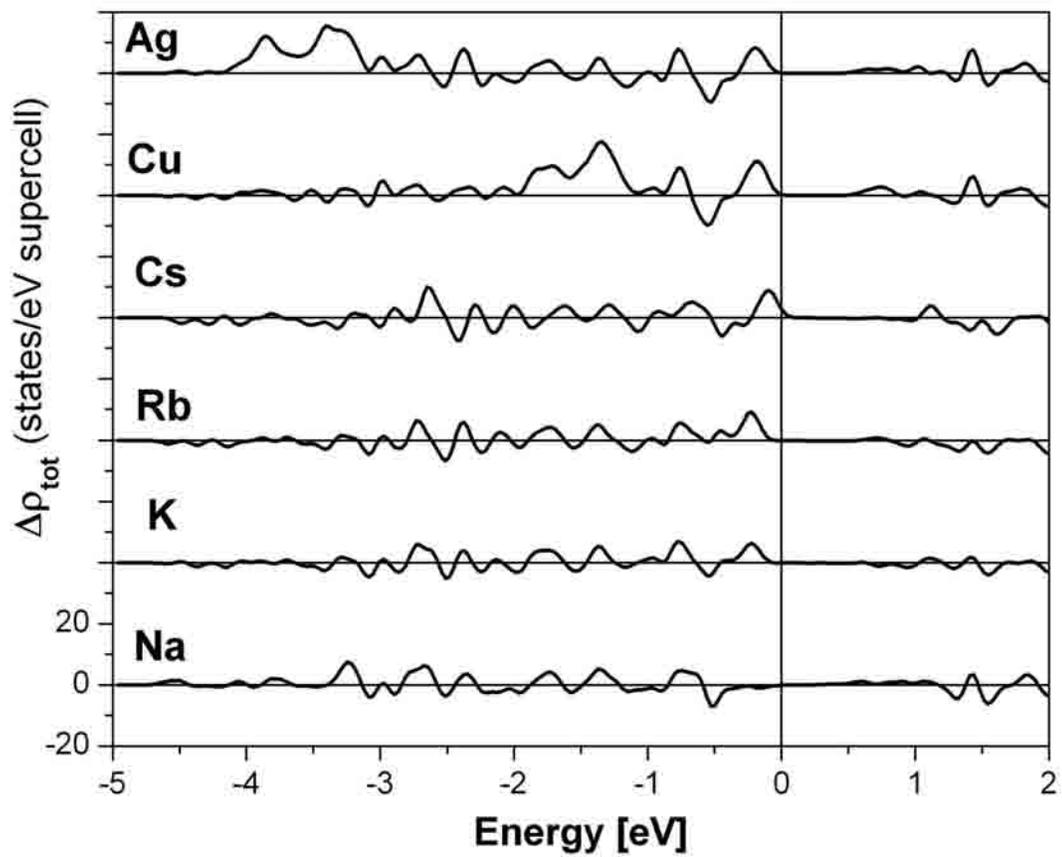

Figure 3

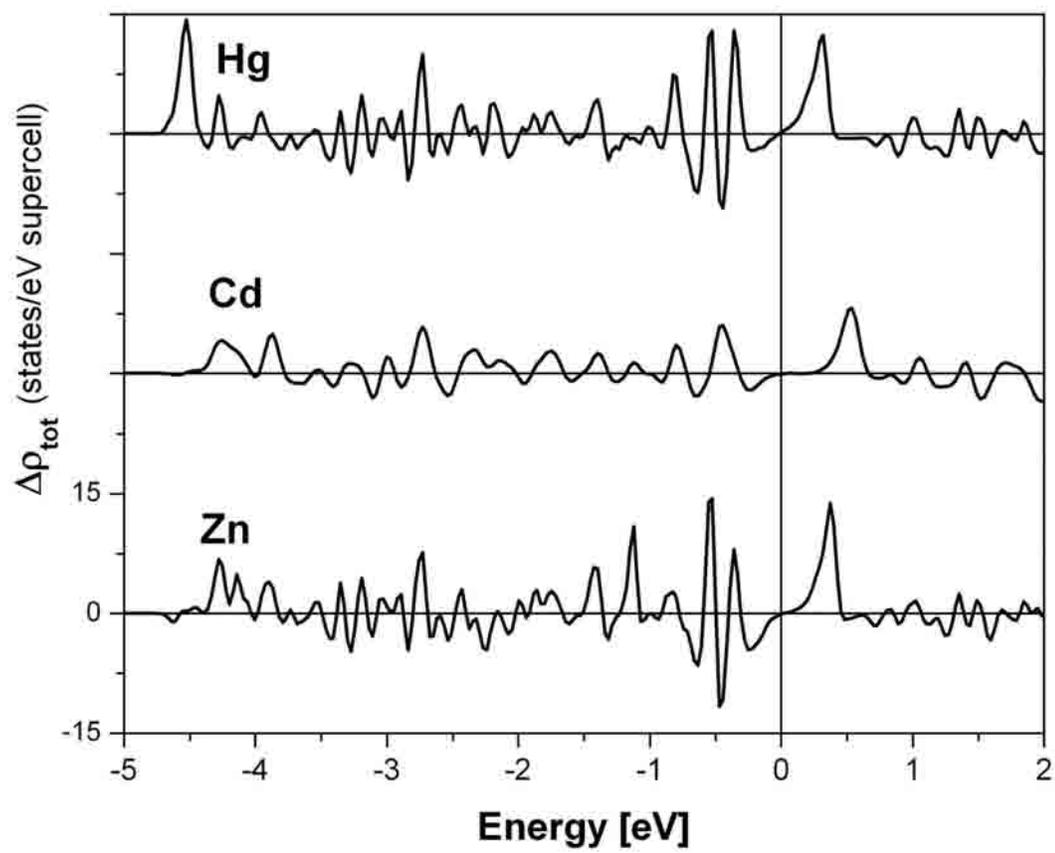

Figure 4

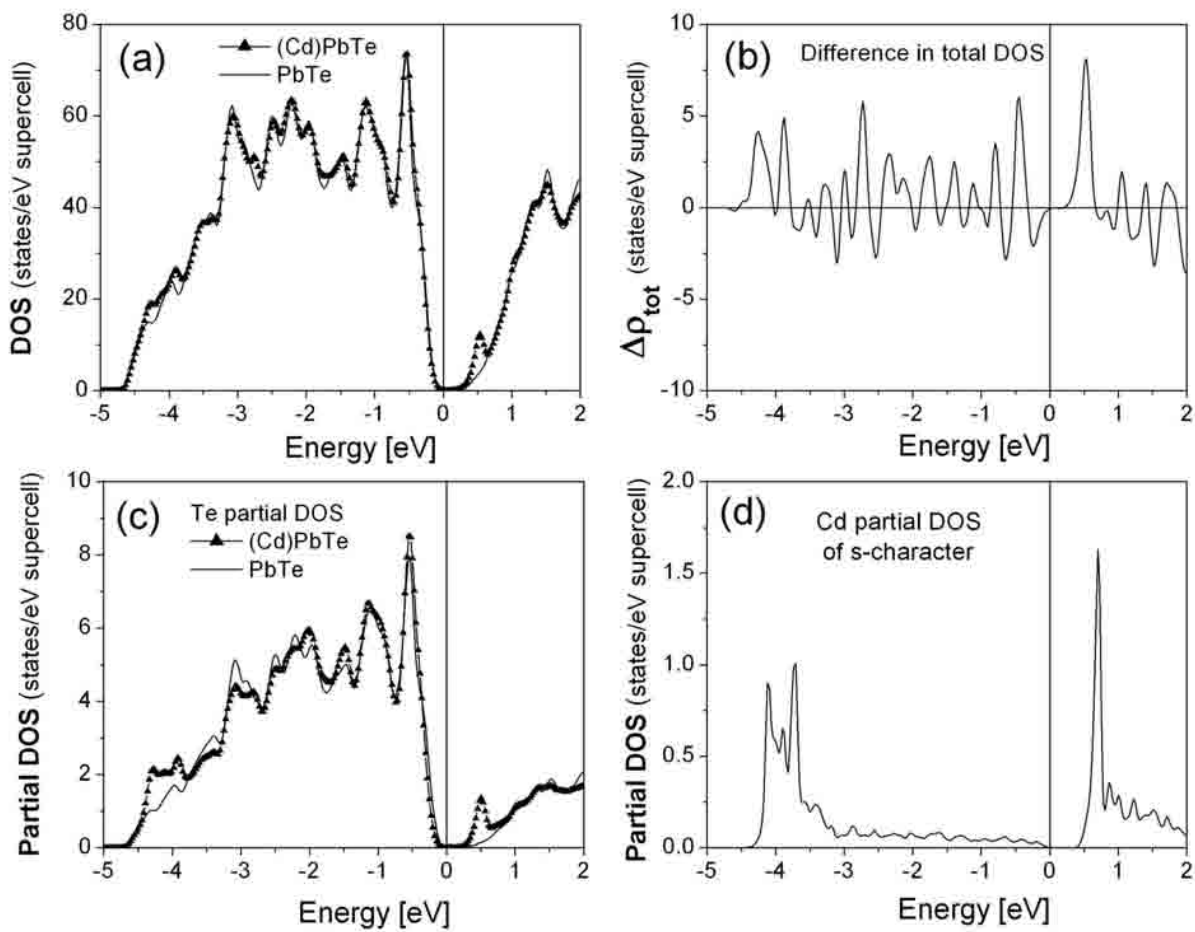

Figure 5

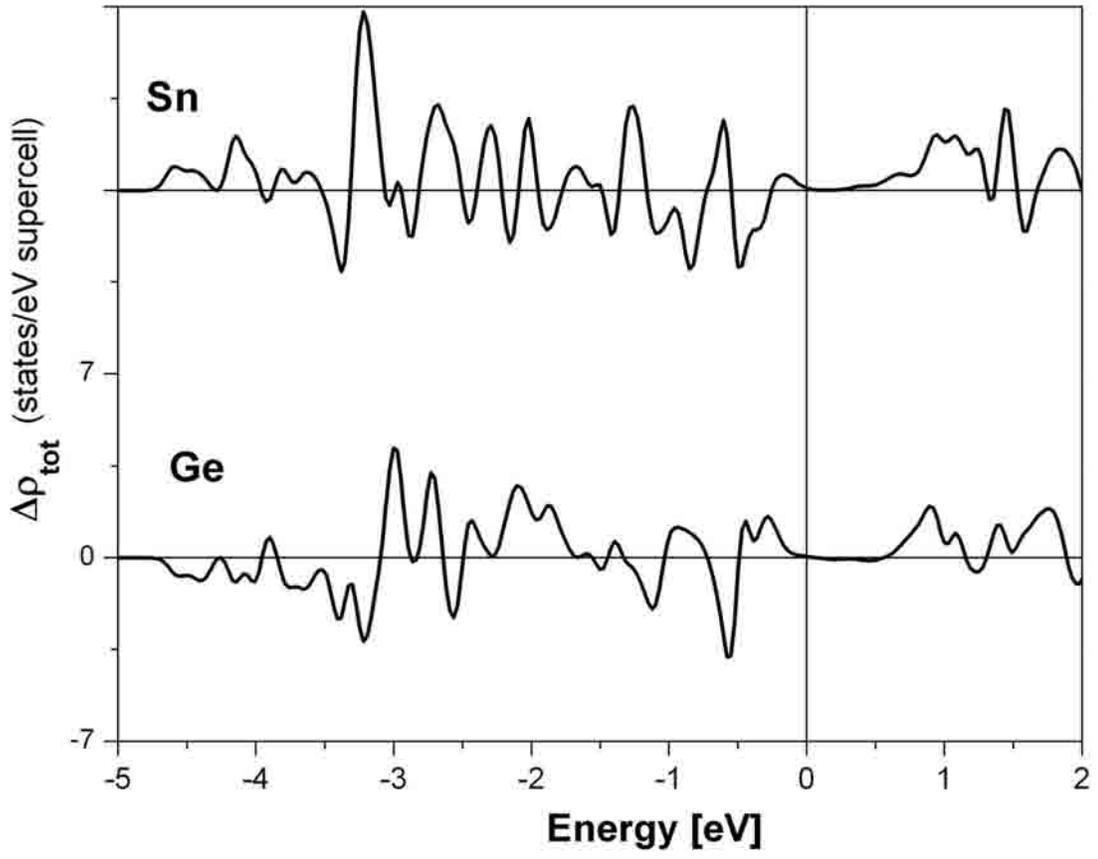

Figure 6

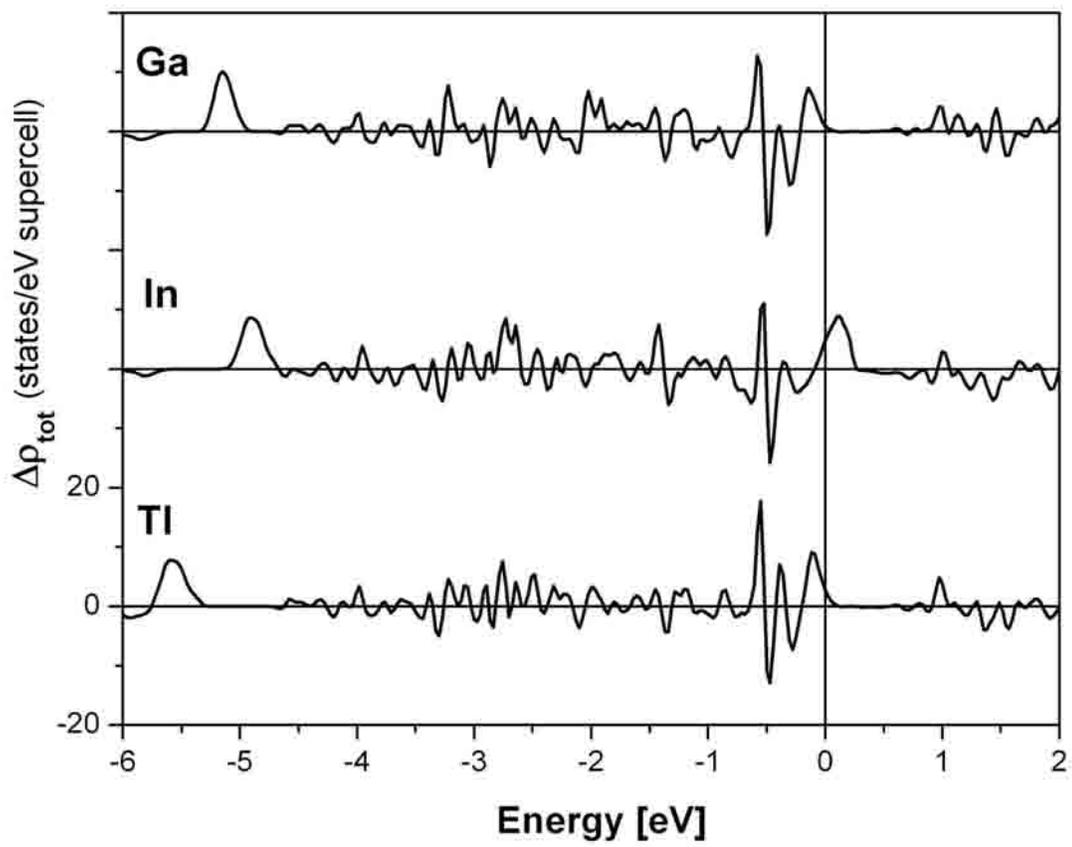

Figure 7

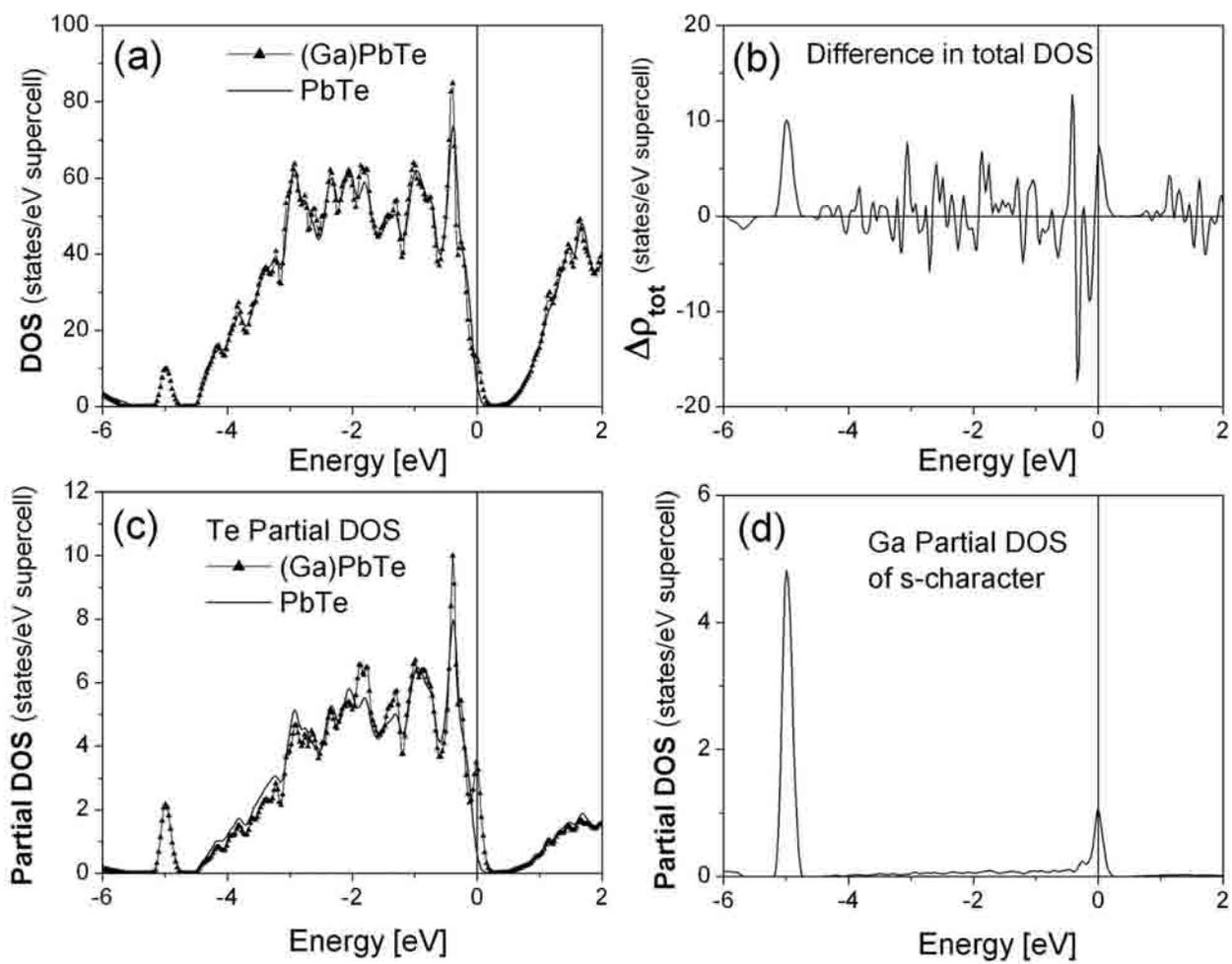

Figure 8

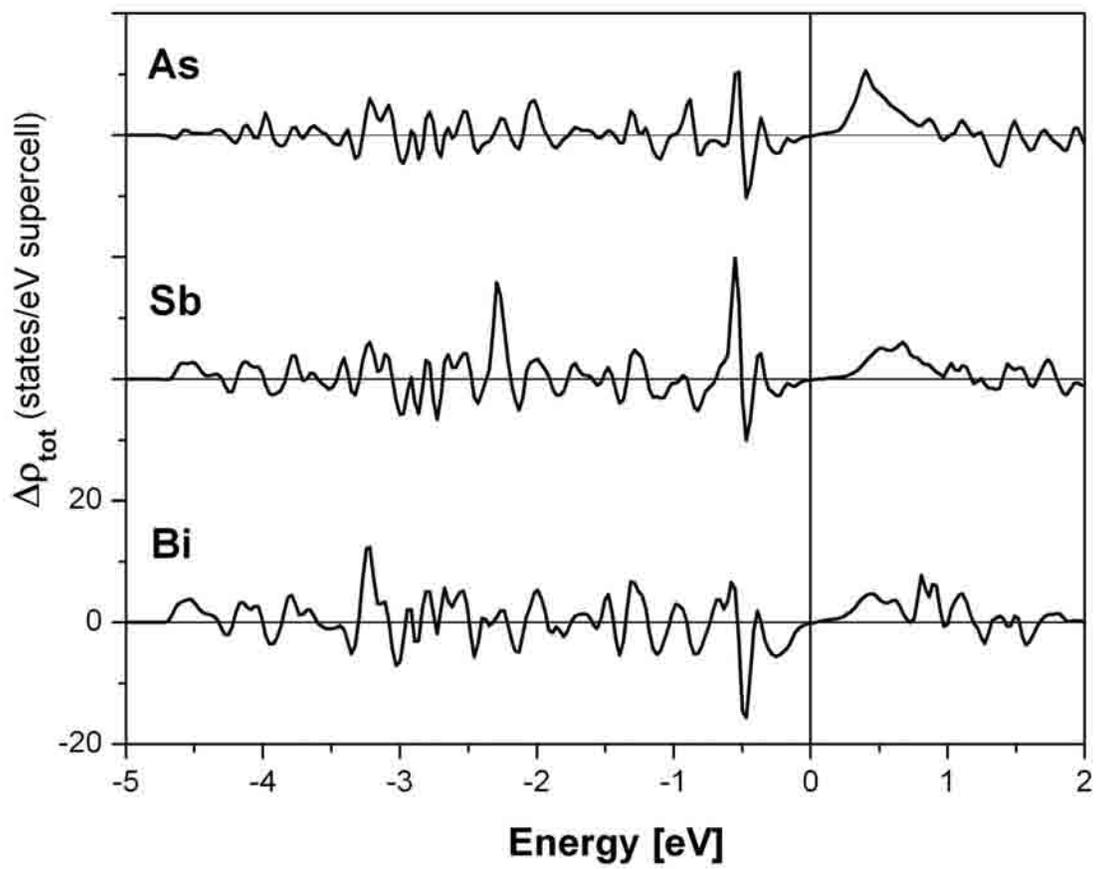

Figure 9

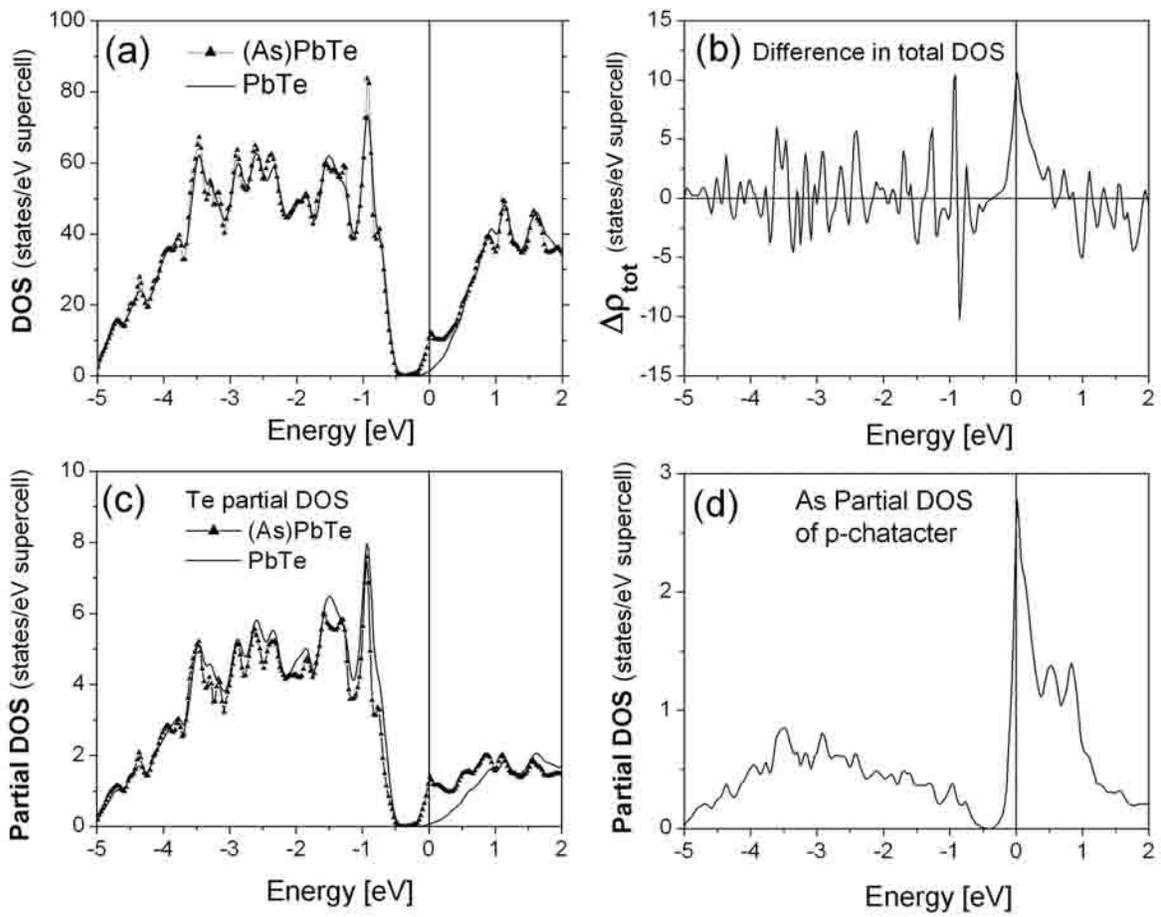

Figure 10

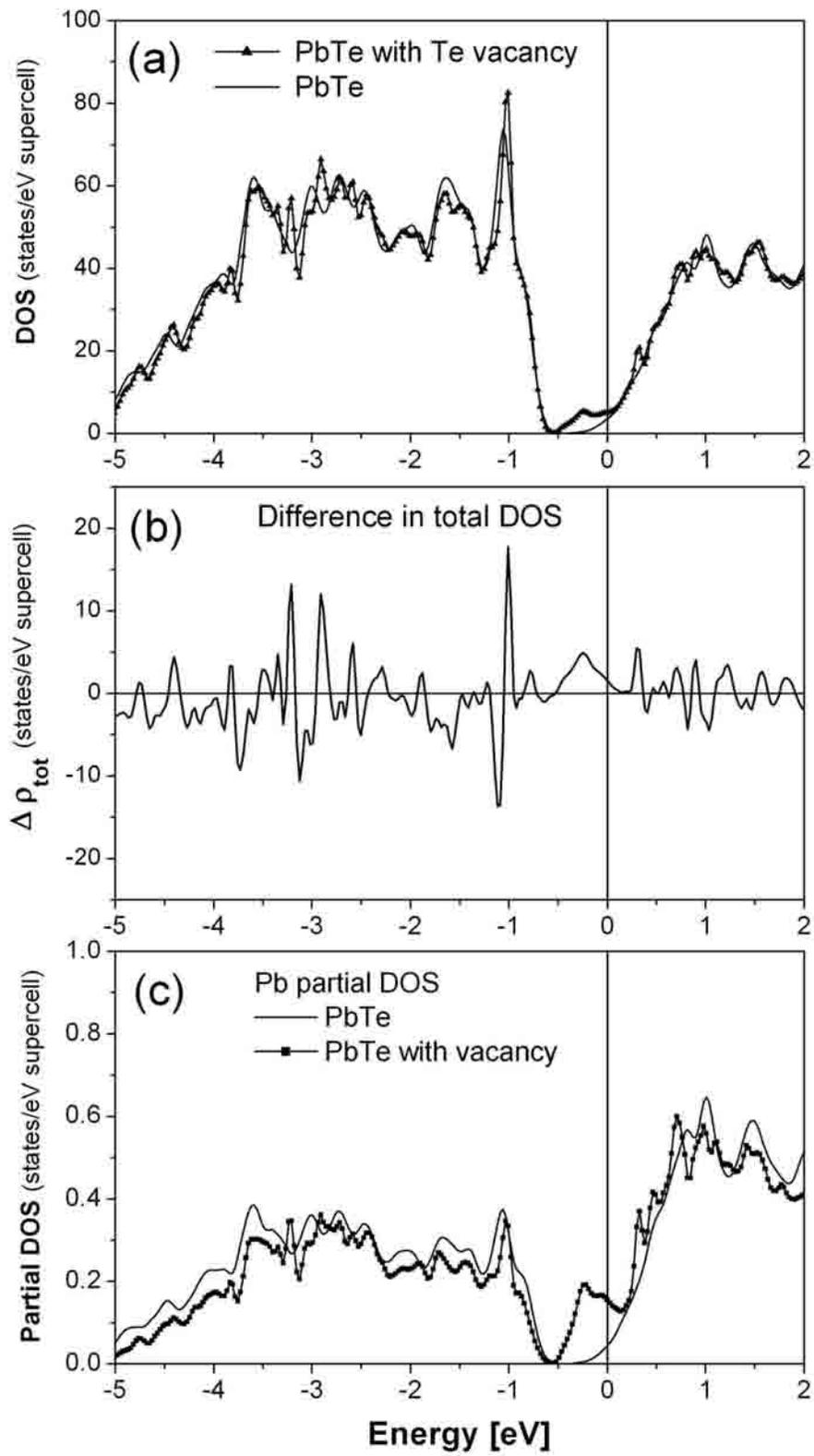

Figure 11

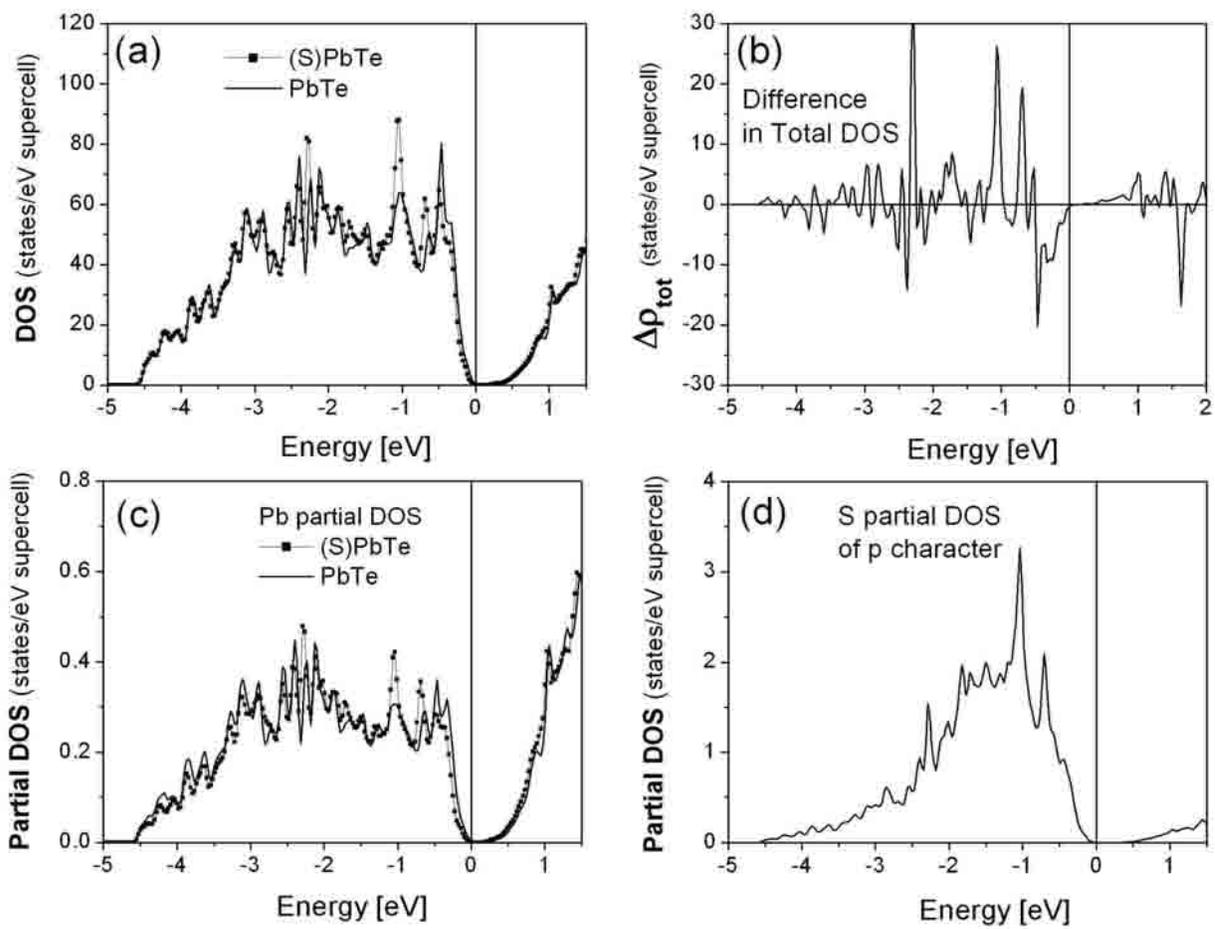

Figure 12

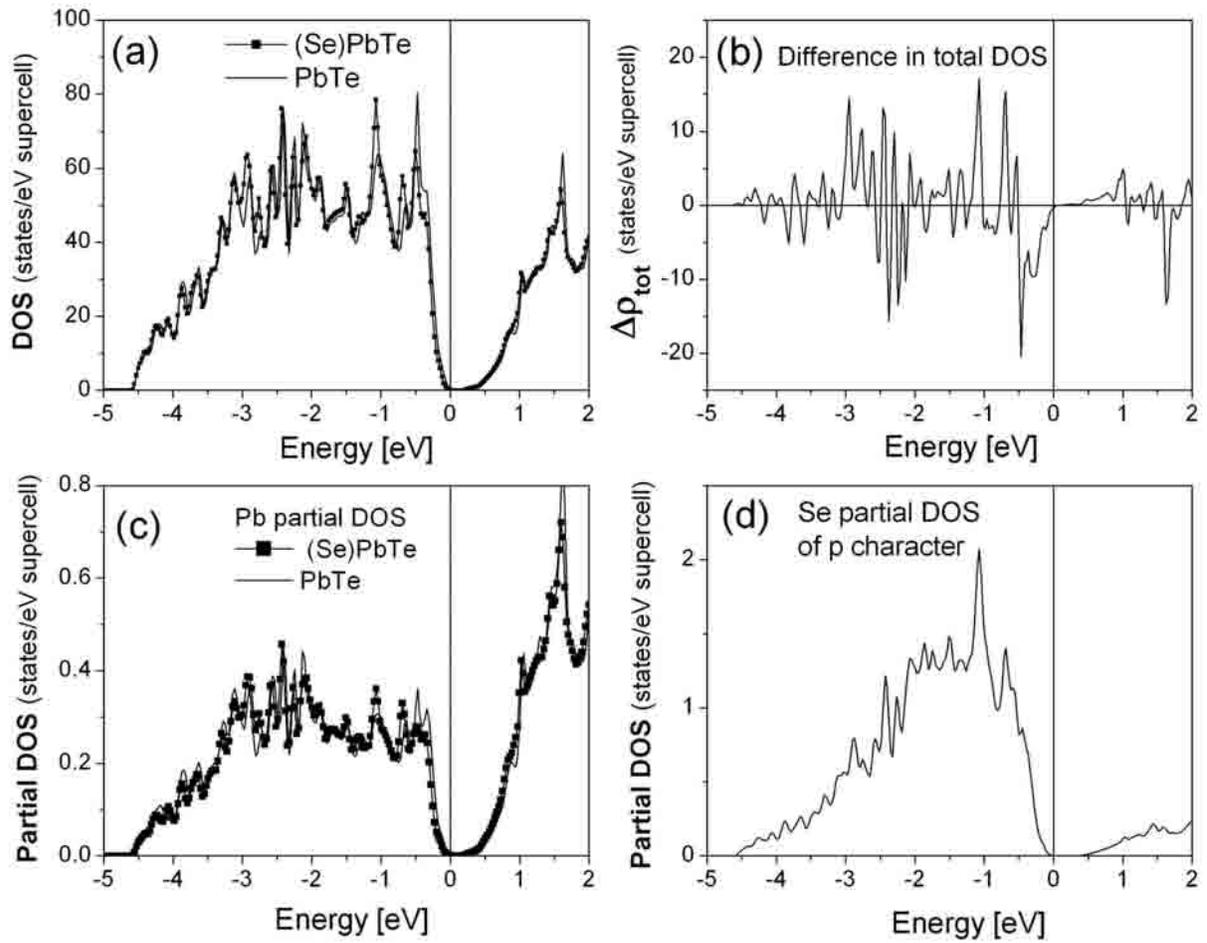

Figure 13

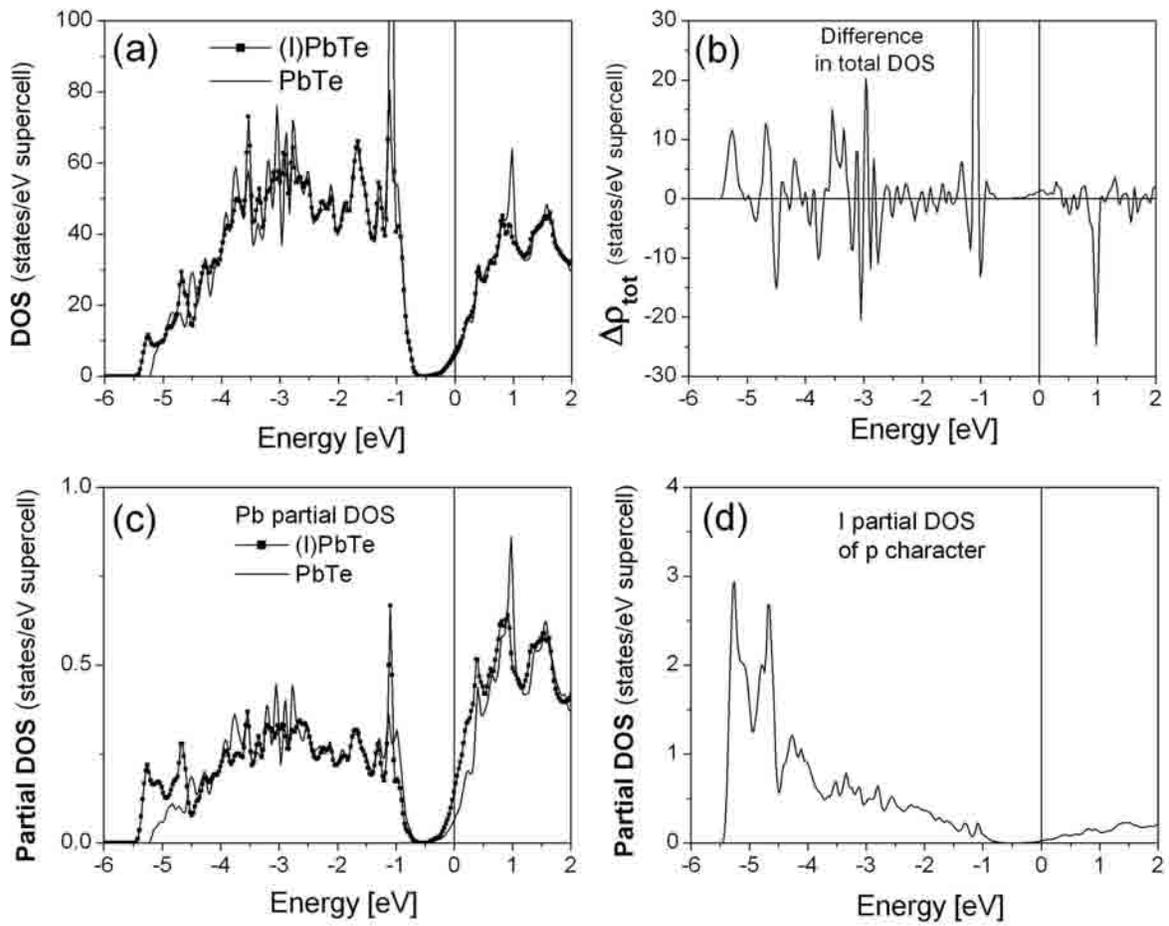

Figure 14